\documentclass[nopreprint,12pt]{elsarticle}

\usepackage{xcolor}
\usepackage{amsmath}
\usepackage{amssymb}
\usepackage{bm}
\usepackage[shortlabels]{enumitem}

 \usepackage[]{subfig}

\usepackage{tikz,pgf} 
\usepackage{lipsum}
\usetikzlibrary{fit}
\usepackage{pgfplotstable}
\pgfplotsset{compat = 1.15}
\usepgfplotslibrary{statistics}
\usetikzlibrary{intersections}
\usepgfplotslibrary{fillbetween}
\usepgfplotslibrary{colorbrewer}
\pgfplotsset{cycle list/OrRd-3}
\usepgfplotslibrary{groupplots}
\usetikzlibrary{plotmarks}
\usetikzlibrary{decorations.pathreplacing}
\usetikzlibrary{spy}

\usepackage[]{textpos}
\usepackage{tikz}
\usetikzlibrary{shapes,arrows}
\usetikzlibrary{calc}
\usepackage[nooldvoltagedirection]{circuitikz}
\usepackage{mathdots}
\usepackage{tabularx}
\usepackage{booktabs}       
\usepackage{siunitx}
\usepackage{makecell}
\DeclareSIUnit\pu{p.u.}
\usepackage{comment}
\usepackage{multirow}
\usepackage{footnote}
\usepackage{pdflscape}
\makesavenoteenv{tabular}
\makesavenoteenv{table}
\usepackage[hidelinks]{hyperref}
\usepackage{dirtytalk}
\usepackage[capitalise]{cleveref}
\usepackage[acronym,toc,nonumberlist]{glossaries} 
\newcommand\norm[2]{\ensuremath{\left\lVert#1\right\rVert_{#2}}}

\newcommand{\dataset}[1]{\ensuremath{\mathcal{D}_{\text{#1}}}}
\newcommand{\NNinput}{\ensuremath{\bm{z}_0}}
\newcommand{\NNoutput}{\ensuremath{\hat{\bm{x}}}}
\newcommand{\Foutput}{\ensuremath{\bm{x}}}
\newcommand{\Ffunction}{\ensuremath{\bm{f}}}

\newcommand{\lossx}{\ensuremath{\mathcal{L}_{x}}}
\newcommand{\lossdt}{\ensuremath{\mathcal{L}_{dt}}}
\newcommand{\lossf}{\ensuremath{\mathcal{L}_{f}}}

\newcommand{\scalelossx}{\ensuremath{\xi_{x, i}}}
\newcommand{\scalelossdt}{\ensuremath{\xi_{dt, i}}}
\newcommand{\scalelossf}{\ensuremath{\xi_{f, i}}}

\newcommand{\weightlossdt}{\ensuremath{\lambda_{dt}}}
\newcommand{\weightlossf}{\ensuremath{\lambda_{f}}}
\newcommand{\weightlossfmax}{\ensuremath{\lambda_{f, \max}}}
\newcommand{\weightlossfzero}{\ensuremath{\lambda_{f,0}}}

\newacronym{NN}{NN}{Neural Network}
\newacronym{PINN}{PINN}{Physics-Informed Neural Network}
\newacronym{BDF}{BDF}{Backward Differentiation Formula}
\newacronym{LBFGS}{L-BFGS}{limited-memory Broyden-Fletcher-Goldfarb-Shanno}
\newacronym{HPC}{HPC}{High Performance Computing}
\newacronym{DAE}{DAE}{Differential-Algebraic Equation}
\newacronym{DTU}{DTU}{Technical University of Denmark}
\newacronym{SciML}{SciML}{Scientific Machine Learning}
\newacronym{ML}{ML}{Machine Learning}
\newacronym{DE}{DE}{Differential Equation}
\newacronym{SAS}{SAS}{Semi-Analytical Solution}
\newacronym{RT}{RT}{Run-Time}
\newacronym{AD}{AD}{Automatic Differentiation}

\definecolor{color_pureNN}{rgb}{0.992, 0.906, 0.78125}
\definecolor{color_dtNN}{rgb}{0.988, 0.730, 0.516}
\definecolor{color_PINN}{rgb}{0.887, 0.289, 0.199}

\journal{Electric Power Systems Research}
\begin{document}




\title{Physics-Informed Neural Networks\\for Time-Domain Simulations:\\Accuracy, Computational Cost, and Flexibility}

\author[inst1]{Jochen Stiasny\corref{cor1}}
\ead{jbest@dtu.dk}
\cortext[cor1]{Corresponding author}


\affiliation[inst1]{organization={Department for Wind and Energy Systems, Technical University of Denmark},
            addressline={Elektrovej}, 
            city={Kgs. Lyngby},
            postcode={2800}, 
            country={Denmark}}

\author[inst1]{Spyros Chatzivasileiadis}
\ead{spchatz@dtu.dk}

\begin{abstract}
The simulation of power system dynamics poses a computationally expensive task. Considering the growing uncertainty of generation and demand patterns, thousands of scenarios need to be continuously assessed to ensure the safety of power systems. \glspl{PINN} have recently emerged as a promising solution for drastically accelerating computations of non-linear dynamical systems. This work investigates the applicability of these methods for power system dynamics, focusing on the dynamic response to load disturbances. Comparing the prediction of \glspl{PINN} to the solution of conventional solvers, we find that \glspl{PINN} can be 10 to 1'000 times faster than conventional solvers. At the same time, we find them to be sufficiently accurate and numerically stable even for large time steps. To facilitate a deeper understanding, this paper also present a new regularisation of \gls{NN} training by introducing a gradient-based term in the loss function. The resulting \glspl{NN}, which we call dtNNs, help us deliver a comprehensive analysis about the strengths and weaknesses of the \gls{NN} based approaches, how incorporating knowledge of the underlying physics affects \gls{NN} performance, and how this compares with conventional solvers for power system dynamics. 
\end{abstract}





\maketitle
\begin{textblock*}{\textwidth}(0.0cm,2.5cm)
\footnotesize
Manuscript published in \@journal\\
DOI: \url{https://doi.org/10.1016/j.epsr.2023.109748}\\
Received 16 March 2023, Revised 7 July 2023, Accepted 3 August 2023\\
© 2023. This manuscript version is made available under the CC-BY-NC-ND 4.0 license \url{https://creativecommons.org/licenses/by-nc-nd/4.0}
\end{textblock*}

\glsresetall

\section{Introduction}\label{sec:introduction}
Time-domain simulations form the backbone in many power system analyses such as transient or voltage stability analyses. However, even the simplest set of governing \glspl{DAE} which can describe the system dynamics sufficiently accurate, can impose a significant computational burden during the analysis. Ways to reduce this computational cost while maintaining a sufficiently high level of accuracy is of paramount importance across all applications in the power systems industry. 

Since, generally speaking, there is no closed form analytical solution for \glspl{DAE} \cite{brenan_numerical_1995}, we revert to numerical methods to approximate the dynamic response. Refs.~\cite{stott_power_1979,sauer_power_1998} provide a good overview on general solution approaches and the modelling in the power system context, and \cite{gurrala_parareal_2016,liu_solving_2020,aristidou_time-domain_2015} summarise important developments, mostly relying on model simplification, decompositions, pre-computing partial solutions, and parallelisations. 

A new avenue to solve ordinary and partial differential equations emerged recently through so-called \Gls{SciML} -- a field, which combines scientific computing with \gls{ML}. \Gls{SciML} has been receiving a lot of attention due to the significant potential speed-ups it can achieve for computationally expensive problems, such as the solution of differential equations. More specifically, the authors in \cite{lagaris_artificial_1998}, already 25 years ago, introduced the idea of using artificial \glspl{NN} to approximate such solutions. The idea is that \glspl{NN} learn from a set of training data to interpolate the solution for data points that lie between the training data with high accuracy. Ref.~\cite{raissi_physics-informed_2018} has revived this effort, now named \glspl{PINN}, which has developed into a growing field within \gls{SciML} as \cite{karniadakis_physics-informed_2021} reviews. The key idea of \glspl{PINN} is to directly incorporate the domain knowledge into the learning process. We do so by evaluating if the \gls{NN} output satisfies the set of \glspl{DAE} during training. If it does not, the parameters of the \gls{NN} are adjusted in the next training iteration until the \gls{NN} output satisfies the DAEs. This approach reduces the need for large training datasets and hence the associated costs for simulating them. Ref.~\cite{misyris_physics-informed_2020} introduced \glspl{PINN} in the field of power systems.

Our ultimate goal is to develop \glspl{PINN} as a solution tool for time-domain simulations in power systems. This paper takes a first step, and identifies the strengths and weaknesses of such a method in comparison with existing solution methods with respect to the application specific requirements on the solution method. Stott elaborated nearly half a century ago that, among others, sufficient accuracy, numerical stability, and flexibility were important characteristics that need to be weighed against the solution speed \cite{stott_power_1979}. In an ideal world, we are looking for tools that are highly accurate, numerically stable, and flexible, and at the same time very fast. Several approaches have been proposed to deal with this trade-off, aiming at being faster (at least during run-time) while maintaining accuracy, numerical stability, and flexibility to the extent possible. Some of the promising ones are based on pre-computing parts of the solution of \glspl{DAE}. For example, \gls{SAS}-methods adopt this approach \cite{gurrala_large_2017,duan_power_2017,wang_timepower_2019}. We can push this idea of pre-computing the solution even further: \glspl{PINN}, and \glspl{NN} in general, pre-compute -- learn -- the entire solution, hence, the computation at run-time is extremely fast. Related works in \cite{moya_dae-pinn_2023,li_machine-learning-based_2020,cui_predicting_2021} introduce alternative \gls{NN} architectures and problem setups, primarily driven by considerations on the achieved accuracy. In contrast, our focus lies on assessing \glspl{PINN} from a perspective of a numerical solution method in which accuracy has to be weighed against other numerical characteristics namely speed, numerical stability and flexibility.  The contributions of this work are the following:

\setlist[enumerate,1]{leftmargin=0.5cm}
\begin{enumerate}
    \item We apply \acrfullpl{PINN} to multi-machine systems and show that \glspl{PINN} can be 10 to 1'000 times faster than conventional methods for time-domain simulations, while achieving sufficient accuracy.
    \item We demonstrate that the trade-off between speed and accuracy for \glspl{PINN}, and \glspl{NN} in general, does not directly relate to power system size but rather to the complexity of the dynamics. Hence, \glspl{NN} can solve larger systems equally fast as small ones, if the complexity of the dynamics is comparable. This is contrary to conventional methods, where the solution time is closely linked to the system size.
    \item We examine further numerical properties of \glspl{NN} for solving \glspl{DAE}. Besides speed, one of their key benefits is that \glspl{NN} do not suffer from numerical instability as they solve without any iterative procedure. We also discuss the challenges of flexibility in different parameter settings and we outline concrete directions for future work to resolve them.
    \item Having shown that \glspl{NN} do have significant benefits and desirable properties, we carry out a comprehensive analysis on the performance and training of \glspl{NN} and \glspl{PINN} that can be helpful for future applications. In this context, we introduce \textit{dtNNs}, a regularised form of \glspl{NN}. dtNNs are an intermediate methodological step between NNs and PINNs as they are regularised by the time derivatives at the training data points. 
\end{enumerate}
\Cref{sec:methodology} describes the construction of a \gls{NN}-based approximation for \glspl{DAE} and how to incorporate physical knowledge in dtNNs and \glspl{PINN}. \Cref{sec:case_study} presents the case study and the training setup. \cref{sec:results} shows the results, on which basis we discuss the route forward in \cref{sec:discussion}. \Cref{sec:conclusion} concludes.

\section{Methodology}\label{sec:methodology}
This section lays out how we train a \gls{NN} that shall be used in time-domain simulations, how the physical equations can be incorporated transforming the \gls{NN} to a dtNN and a \gls{PINN}, and how the resulting approximation is assessed.

\subsection{Approximating the solution to a dynamical system}
A dynamical system is characterised by its temporal evolution being dependent on the system's state variables $\bm{x}$, the algebraic variables $\bm{y}$ and the control inputs $\bm{u}$:
\begin{subequations}
\begin{align}
    \frac{d}{dt}\bm{x} &= \bm{f}_{\text{DAE}}\left(\bm{x}(t), \bm{y}(t), \bm{u}\right)\label{eq:f_DAE}\\
     \bm{0} &= \bm{g}_{\text{DAE}}\left(\bm{x}(t), \bm{y}(t), \bm{u}\right)\label{eq:g_DAE}.
\end{align}
\end{subequations}
For clarity and ease of implementation, we express \labelcref{eq:f_DAE,eq:g_DAE} as
\begin{align}
        \bm{M}\frac{d}{dt}\bm{x} &= \bm{f}(\bm{x}(t), \bm{u}).\label{eq:dynamical_system}
\end{align}
by incorporating $\bm{y}$ into $\bm{x}$ and adding $\bm{M}$, which is a diagonal matrix to distinguish if a state $x_i$ is differential $(M_{ii} \neq 0)$ or algebraic $(M_{ii} = 0)$. We will use a \gls{NN} to define an explicit function $\hat{\bm{x}}(t)$ that shall approximate the solution $\bm{x}(t)$ for all $t \in [t_0, t^{\max}]$, i.e., for the entire \textit{trajectory}, starting from the initial condition $\bm{x}(t_0) = \bm{x}_0$.

\subsection{Neural network as function approximator}
We use a standard feed-forward \gls{NN} with $K$ hidden layers that implements a sequence of linear combinations and non-linear activation functions $\sigma(\cdot)$. In theory, a \gls{NN} with a single hidden layer already constitutes a universal function approximator \cite{cybenko_approximation_1989} if it is wide enough, i.e., the hidden layer consists of enough neurons $N_K$. In practice, restrictions on the width and the process of determining the \gls{NN}'s parameters might limit this universality as \cite{goodfellow_deep_2016} elaborates. Still, a multi-layer \gls{NN} in the form of \eqref{eq:NN_equations} provides us with a powerful function approximator:
\begin{subequations}\label{eq:NN_equations}
\begin{alignat}{2}
    [t, \bm{x}_0^\top, \bm{u}^\top]^\top &= \bm{z}_0 &&\label{eq:NN_input}\\
    \bm{z}_{k+1} &= \sigma{(\bm{W}_{k+1} \bm{z}_k + \bm{b}_{k+1})} && \quad \forall k = 0, 1, ..., K-1\label{eq:NN_hidden_layers}\\
    \hat{\bm{x}} &= \bm{W}_{K+1} \bm{z}_K + \bm{b}_{K+1}.&&\label{eq:NN_output}
\end{alignat}
\end{subequations}
The NN output \NNoutput{} is the system state at the prediction time $t$. The input \NNinput{} is composed of the prediction time $t$, the initial condition $\bm{x}_0$ and the control input $\bm{u}$. The weight matrices $\bm{W}_k$ and bias vectors $\bm{b}_i$ form the adjustable parameters $\bm{\theta}$ of the \gls{NN}.

For the training process, we compile a training dataset \dataset{train}, that maps $\NNinput \mapsto \Foutput$ for a chosen input domain $\mathcal{Z}$ and contains $N = |\dataset{train}|$ points. For our purposes, the input domain is a discrete set of the prediction time, e.g. from \SI{0}{\second} until \SI{10}{\second} with a step size of \SI{0.2}{\second}, and a set of different initial conditions and control inputs, e.g. different power disturbances. The output domain is the rotor angle and frequency at each of the prediction time steps and for each of the studied disturbances.
\begin{align}
    \dataset{train}: \NNinput \mapsto \Foutput \qquad \NNinput \in \mathcal{Z}. \label{eq:dataset_definition}
\end{align}
During training we adjust the \gls{NN}'s parameters $\bm{\theta}$ with an iterative gradient-based optimisation algorithm to minimise the so-called \textit{loss} $\mathcal{L}$ for $\mathcal{D}_{train}$
\begin{subequations}\label{eq:NN_optimisation}
\begin{align}
    \min_{\bm{\theta}} \quad &\mathcal{L}(\dataset{train})\\
    \text{s.t.} \quad & \eqref{eq:NN_input} - \eqref{eq:NN_output}.
\end{align}
\end{subequations}
We do not aim for optimality of \labelcref{eq:NN_optimisation} -- this would lead to over-fitting -- but rather search for parameters $\bm{\theta}$ (i.e., values of the weights and biases) and hyper-parameters (e.g., number of layers $K$ and neurons per layer $N_K$) that yield a low generalisation error $\mathcal{L}(\dataset{test})$ which we assess on a separate test dataset \dataset{test}. During training we use a validation dataset \dataset{validation} to obtain an estimate of the generalisation error, so that the final evaluation with \dataset{test} remains unbiased. It is important, that all three datasets stem from the same sampling procedure and input domain $\mathcal{Z}$.

\subsection{Loss function and regularisation: \glspl{NN}, dtNNs, and \glspl{PINN}}

\subsubsection{Loss Function for Neural Networks}
The simplest loss function for such a problem is to define the loss as the mismatch between the \gls{NN} prediction \NNoutput{} and the ground truth or target \Foutput{}, and measure it using the L2-norm. To account for different orders of magnitude (for example, the voltage angles in radians are often much larger than frequency deviations expressed in \si{\pu}) and levels of variations of the individual states $\bm{x}$, we first apply a scaling factor \scalelossx{} to the error computed per state $i$. A physics-agnostic choice of \scalelossx{} could be to use the state's standard deviation in the training dataset; for more details please see \cref{subsec:NN_setup}. We then apply the squared L2-norm for each data point $j$ and take the average across the dataset \dataset{} to obtain the loss \lossx{}
\begin{align}
    \lossx(\dataset{}) &= \frac{1}{|\dataset{}|}\sum_{j=1}^{|\dataset{}|} \norm{\left(\frac{\hat{x}_i^j- x_i^j}{\scalelossx{}}\right)}{2}^2\label{eq:loss_data}.
\end{align}

\subsubsection{dtNNs}
As an intermediate step between standard \glspl{NN} and \glspl{PINN}, in this subsection we introduce a new regularisation term to loss function \eqref{eq:loss_data}. We do so to avoid the previously mentioned over-fitting and improve the generalisation performance of the \glspl{NN}. To the best of our knowledge, this paper is the first to introduce a regularisation term based on the update function \Ffunction{}(\Foutput{}) from \eqref{eq:dynamical_system}. Using the tool of \gls{AD} \cite{baydin_automatic_2018}, we can compute the derivative of the \gls{NN}, i.e., the time derivative of the approximated trajectory, $\frac{d}{dt}\NNoutput{}$ and compute a loss analogous to \eqref{eq:loss_data} (with a scaling factor \scalelossdt{}):
\begin{align}
    \lossdt(\dataset{}) &= \frac{1}{|\dataset{}|}\sum_{j=1}^{|\dataset{}|} \norm{\left(\frac{\frac{d}{dt}\hat{x}_i^j- \frac{d}{dt} x_i^j}{\scalelossdt{}}\right)}{2}^2\label{eq:loss_dtNN}
\end{align}

\subsubsection{PINNs}
As \cite{lagaris_artificial_1998,raissi_physics-informed_2018} introduced generally, and \cite{misyris_physics-informed_2020} for power systems, we can also regularise such a \gls{NN} by comparing the derivative of the \gls{NN} $\frac{d}{dt}\NNoutput{}$ with the update function evaluated based on the estimated state \Ffunction{}(\NNoutput{}):
\begin{align}
    \lossf(\mathcal{D}_f) &= \frac{1}{|\mathcal{D}_f|}\sum_{j=1}^{|\mathcal{D}_f|} \norm{\left(\frac{M_{ii}\frac{d}{dt}\hat{x}_i^j - f_i(\hat{\bm{x}}^j)}{\scalelossf{}}\right)}{2}^2\label{eq:loss_PINN}
\end{align}
This physics-loss does not require the ground truth state \Foutput{} or its derivative. Quite the contrary, this loss can be queried for any desired point without requiring any form of simulation. We therefore can evaluate a dataset $\mathcal{D}_f$ of randomly sampled or ordered \textit{collocation points} that map to 0
\begin{align}
    \mathcal{D}_f: \NNinput \mapsto \bm{0} \qquad \NNinput \in \mathcal{Z}. \label{eq:collocation_definition}
\end{align}
to essentially assess how well the \gls{NN} approximation follows the physics - any point where this physics loss equals zero is in line with the governing physics of \eqref{eq:dynamical_system}. However, \eqref{eq:collocation_definition} defines a mapping that is not bijective, hence, $\mathcal{L}_{f}(\mathcal{D}_f) = 0$ does not imply that the desired trajectory is perfectly matched, only that a trajectory complying with \eqref{eq:dynamical_system} is matched. As an example, an exact prediction of the steady state of the system will yield $\mathcal{L}_{f}(\mathcal{D}_f) = 0$ even though the target trajectory in \dataset{train} is different.

\subsubsection{Combined loss function during training}
To obtain a single objective or loss value for the training problem \eqref{eq:NN_optimisation}, we weigh the three terms as follows:
\begin{align}
    \mathcal{L} &= \lossx + \weightlossdt \lossdt + \weightlossf \lossf,
\end{align}
where \weightlossdt{} and \weightlossf{} are hyper-parameters of the problem. Subsequently, we refer to a \gls{NN} trained with $\weightlossdt{} = 0, \weightlossf{} = 0$ as \say{vanilla NN}\footnote{In \cite{hastie_elements_2009} \say{vanilla \gls{NN}} refers to a feed-forward \gls{NN} with a single layer, we adopt the term nonetheless for clarity as it expresses the idea of a \gls{NN} without any regularisation well.}, with $\weightlossdt{} \neq 0$, $\weightlossf{} = 0$ as \say{dtNN}, and with $\weightlossdt{} \neq 0, \weightlossf{} \neq 0$ as \say{\gls{PINN}}.

\subsection{Accuracy metrics}

To compare across the different methods and setups, we monitor the loss $\mathcal{L}_x$ in \eqref{eq:loss_data} as the comparison metric throughout the training and evaluation process and as an accuracy metric for the performance assessment. To get a more detailed picture, we also consider the loss value of single points, i.e., before calculating the mean in \labelcref{eq:loss_data}.
However, the loss is dependent on the chosen values for $\scalelossx{}$ and does not provide an easily interpretable meaning. Therefore, we use the maximum absolute error 
\begin{align}
    \max AE_{\mathcal{S}} &= \max_{i \in \mathcal{S}, j \in \dataset{test}}\left(\left|\hat{x}_i^j- x_i^j\right|\right)\label{eq:max_abs_error}
\end{align}
as an additional metric for assessment purposes, i.e., based on \dataset{test}, but not during training. Whereas a state-by-state metric would capture most details, we opt to compute the maximum absolute error across meaningful groups of states $i \in \mathcal{S}$ that are of the same units and magnitudes. This aligns with the engineering perspective on the desired accuracy of a method.

\section{Case study}\label{sec:case_study}
This section introduces the test cases and the details of the NN training.

\subsection{Power system - Kundur 11-bus and IEEE 39-bus system}\label{subsec:test_cases}

As a study setup, we investigate the dynamic response of a power system to a load disturbance. We use a second order model to represent each of the generators in the system. The update equation \labelcref{eq:dynamical_system} formulates for generator buses as
\begin{align}
    \begin{bmatrix} 1 & 0 \\ 0 & 2 H_i \omega_0 \end{bmatrix} \frac{d}{dt} \begin{bmatrix} \delta_{i} \\ \Delta \omega_i \end{bmatrix} &= \begin{bmatrix}
    \Delta \omega_i \\
    P_{mech,i} - D_i\Delta \omega_i + P_{e,i}
    \end{bmatrix}
    \intertext{and for load buses as}
    \begin{bmatrix}d_i \omega_0\end{bmatrix} \frac{d}{dt} \begin{bmatrix}
    \delta_{i}
    \end{bmatrix} &= \begin{bmatrix}
    P_{mech,i} + P_{e,i}
    \end{bmatrix}
\end{align}
where $P_{mech,i} = P_{set, i} + P_{dist, i}$ at bus $i$, with $P_{set, i}$ representing the power setpoint and $P_{dist, i}$ the disturbance . The states $\bm{x}$ are the bus voltage angle $\delta_i$ and the frequency deviation $\Delta \omega_i$ for generator buses, and the bus voltage angle $\delta_i$ for the load buses. The buses are linked through the active power flows in the network defined by the admittance matrix $\Bar{\bm{Y}}_{bus}$ and the vector of complex voltages $\Bar{\bm{V}} = \bm{V}_{m} e^{j\bm{\delta}}$, where the vector $\bm{V}_m$ collects the voltage magnitudes and $\bm{\delta}$ the bus voltage angles:
\begin{align}
    \bm{P}_{e} &= \Re \left(\Bar{\bm{V}} (\Bar{\bm{Y}}_{bus} \Bar{\bm{V}})^*\right).
\end{align}
The $*$ indicates the complex conjugate and $P_{e,i}$ corresponds to the $i$-th entry of vector $\bm{P}_{e}$, i.e., the active power balance at bus $i$. 
In \cref{sec:results}, we demonstrate the methodology on the Kundur 2-area system (11 buses, 4 generators) and the IEEE 39-bus test system (39 buses, 10 generators). For both systems we are using the base power of $\SI{100}{MVA}$ and $\omega_0 = \SI{60}{\hertz}$. The network parameters and set-points stem from the case description of Kundur \cite[p.~813]{kundur_power_1994} and the IEEE 39-bus test case in Matpower \cite{zimmerman_matpower_2011}. The values for the inertia of the generators $H_i$ are [6.5, 6.5, 6.175, 6.175] \si{\pu} for the 11-bus case and [500.0, 30.3, 35.8, 38.6, 26.0, 34.8, 26.4, 24.3, 34.5, 42.0] \si{\pu} for the 39-bus case. The damping factor was set to $D_i = 0.05 \frac{\omega_0}{P_{set,i}}$ in both cases and for the loads to $d_i = 1.0 \frac{P_{set,i}}{\omega_0}$ and $d_i = 0.2 \frac{P_{set,i}}{\omega_0}$ respectively.

\subsection{NN training implementation}\label{subsec:NN_setup}

The entire workflow is implemented in Python 3.8 and available under \cite{stiasny_publicly_2022}. When we use the conventional numerical approaches to carry out the time-domain simulations for this system, the dynamical system is simulated using the Assimulo package \cite{andersson_assimulo_2015} which implements various solution methods for systems of \glspl{DAE}. The training process utilises PyTorch \cite{paszke_pytorch_2019} for the learning process and WandB \cite{biewald_experiment_2020} for monitoring and processing the workflow. The implementation builds on \cite{stiasny_closing_2022} for the steps of the workflow.

All datasets comprise the simulated response of the system over a period of \SI{20}{\second} to a disturbance. The tested disturbance is the step response to an instantaneous loss of load $|P_{dist,i}|$ at bus $i$ with a magnitude between \SI{0}{\pu} and \SI{10}{\pu}, where $i=7$ for the 11-bus system and $i=20$ for the 39-bus system. We record these data in increments of $\Delta t$ and $\Delta P$. The test dataset \dataset{test} which shall serve as a ground truth uses $\Delta t = \SI{0.05}{\second}$ and $\Delta P = \SI{0.05}{\pu}$, resulting in $|\dataset{test}| = 401 \times 201 = 80601$ points. This large size of \dataset{test} is chosen as it densely covers the input domain, hence, we obtain a reliable estimate of the maximum absolute error \labelcref{eq:max_abs_error}. For the training datasets \dataset{train} used in \cref{subsec:NN_training} we create datasets with $\Delta t \in [\num{0.2}, \num{1.0}, \num{2.0}]\si{\second}$ and $\Delta P \in [\num{0.2}, \num{1.0}, \num{2.0}]\si{\pu}$. The validation datasets \dataset{validation} for those scenarios are offset by $\frac{\Delta t}{2}$ and $\frac{\Delta P}{2}$.

For the scalings \scalelossx{} in \eqref{eq:loss_data}, we calculate the average standard deviation $\sigma$ across all voltage angle differences $\delta_{ij}$\footnote{The training process benefits from using the voltage angle difference $\delta_{ij} = \delta_i - \delta_j$, where $j$ indicates a reference bus, as the output of the \gls{NN}. The prediction becomes easier as the occurring drift in the dataset with respect to the variable $t$ is significantly reduced.} and all frequency deviations $\Delta \omega_i$, here the relevant groups of states $\mathcal{S}$:
\begin{align}
    \scalelossx{} &= \frac{1}{|\mathcal{S}|}\sum_{i \in \mathcal{S}} \sigma (x_i(\dataset{}))
\end{align}
Thereby, we aim for equal levels of error within all $\delta_{ij}$ and $\Delta \omega_i$ states and account for the difference in magnitude between them.  \scalelossdt{} and \scalelossf{} are all set to 1.0 to avoid adding further hyperparameters, more elaborate choices based on system analysis or the database are conceivable. During training and testing \scalelossx{} is based on \dataset{train} and \dataset{test} respectively.

The regularisation weights \weightlossdt{} and \weightlossf{} are hyperparameters. For the latter, we incorporate a fade-in dependent on the current epoch $E$ :
\begin{align}
    \weightlossf{}(E) = \min{\left(\weightlossfmax{}; \weightlossfzero{} \; 10^{E/E'}\right)},
\end{align}
where \weightlossfmax{} is the maximum and \weightlossfzero{} the initial regularisation weight and $E'$ determines the ``speed'' of the fade-in. The fade-in causes that \lossx{} and \lossdt{} are first minimised and then \lossf{} helps for ``fine-tuning'' and better generalisation. We apply the \gls{LBFGS} algorithm implemented in PyTorch in the training process, a standard optimiser for PINNs as \cite{cuomo_scientific_2022} reviews. The set of hyperparameters comprises $K$, $N_K$, \weightlossdt{}, \weightlossfmax{}, \weightlossfzero{} $E'$, and additional \gls{LBFGS} parameters. \Cref{tbl:hyperparameters_accuracy} reports the used hyperparameters for the scenarios (letters A-E) in \cref{subsec:NN_training}, \cref{subsec:NN_run_time} is based on scenario E for vanilla NNs.
\begin{table}[!ht]
  \caption{Overview of the tuning range and the selected values of the involved hyperparameters.}
  \label{tbl:hyperparameters_accuracy}
  \centering
  \resizebox{\columnwidth}{!}{%
  \renewcommand{\arraystretch}{1.2}
  \begin{tabular}{lccccccc}
    \toprule
     &  & \multicolumn{2}{c}{vanilla NN} &\multicolumn{2}{c}{dtNN} & \multicolumn{2}{c}{PINN}\\
    Hyper-parameter & Tuning range & A-D & E & A-D & E & A-D & E\\
    \midrule
    Number of layers $K$ & [2, 3, 4, 5] & 5 & 5 & 5 & 5 & 5 & 5 \\
    Nodes per layer $N_K$ & [16, 32, 64, 128] & 32 & 32 & 32 & 32 & 32 & 32 \\
    \midrule
    dt regularisation \weightlossdt & [0.01 - 2.0] & - & - & 0.3 & 1.0 & 0.01 & 0.01\\
    Physics regularisation \weightlossfmax & [0.005 - 10] & - & - & - & - & 0.5 & 0.01\\
    Fade-in speed $E'$ & [10 - 50] & - & - & - & - & 15 & 15\\
    \midrule
    Initial learning rate & [0.1 - 2.0] & 1.0 & 1.6 & 0.5 & 2.0 & 1.2 & 1.0\\
    History size & [100 - 150] & 140 & 140 & 120 & 120 & 120 & 120 \\
    Maximum iterations & [18 - 28] & 22 & 22 & 23 & 20 & 20 & 19\\
    \bottomrule
  \end{tabular}
  }
\end{table}

The hyperbolic tangent ($\tanh{}$) is selected as the activation function $\sigma$ as it is continuously differentiable. We initialise the \gls{NN} weights and biases with samples from the distribution described in \cite{glorot_understanding_2010} and achieve different initial values by altering the seed of the random number generator. All training and timing was performed on the \gls{HPC} cluster at the \gls{DTU} with nodes of 2xIntel Xeon Processor 2650v4 (12 core, 2.20GHz) and 256 GB memory of which we used 4 cores per training run.

\section{Results}\label{sec:results}
We first show in this section an assessment of \glspl{NN} at run-time that highlights their methodological advantages compared to conventional solvers. We then perform a comprehensive analysis of the required training phase and the effect of physics regularisation.

\subsection{\glspl{NN} at run-time - opportunities for accuracy and computational cost}\label{subsec:NN_run_time}

\begin{figure}[!th]
    \centering
    \includegraphics{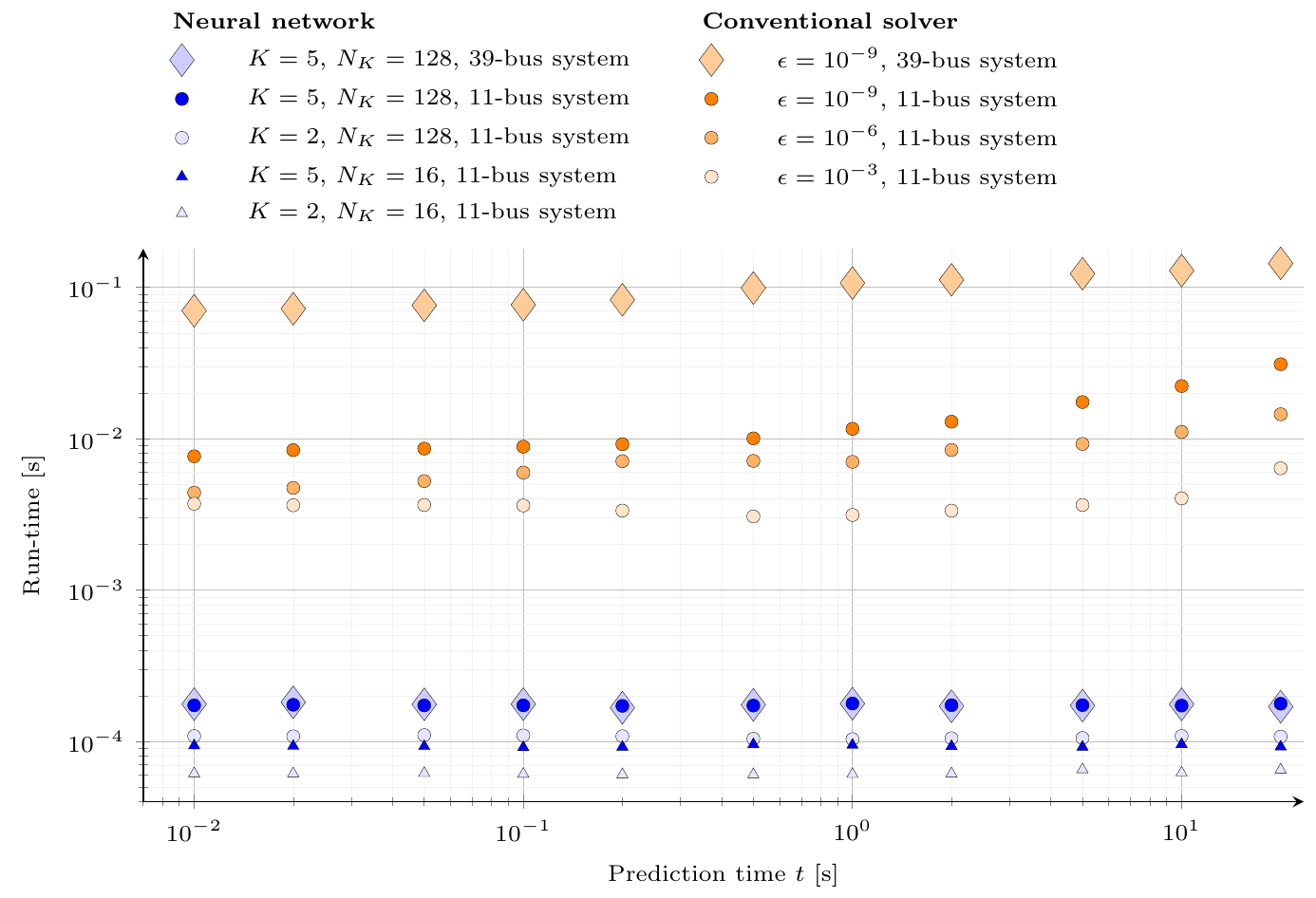}
    \caption{Run-time as a function of the prediction time $t$ for \glspl{NN} of different size and a conventional solver with varied tolerance settings $\epsilon$. Tests for the 11-bus and 39-bus system with a disturbance $P_i = \SI{6.09}{\pu}$.}
    \label{fig:runtime_aspects}
\end{figure}

The primary motivation for the use of \gls{NN}-based solution approaches is their extremely fast evaluation. \Cref{fig:runtime_aspects} shows the run-time for different prediction times. The \glspl{NN} return the value of the states at prediction time $t$ between 10 and 1'000 times faster than the conventional solvers depending on three factors: the prediction time, the power system size, and the solver/\gls{NN} settings.

First, for \glspl{NN} the run-time is independent of the prediction time as the prediction only requires a single evaluation of the \gls{NN}. In contrast, the conventional solver's run-time increases with larger prediction times as more internal time steps are required. Second, the power system size strongly affects the conventional solver's run-time as shown by the increase when moving from the 11-bus to the 39-bus system. For the \gls{NN}, it causes only a negligible change in run-time as only the last layer of the \gls{NN} changes in size according to the number of states of the system, see \labelcref{eq:NN_output}. Third, the \say{solver settings} play an important role; for conventional solvers, the internal tolerance setting $\epsilon$ governs its evaluation speed, while for the \gls{NN} the size, i.e., its number of layers $K$ and number of neurons per layer $N_K$, determine the run-time.

\begin{figure}[!th]
    \centering
    \includegraphics{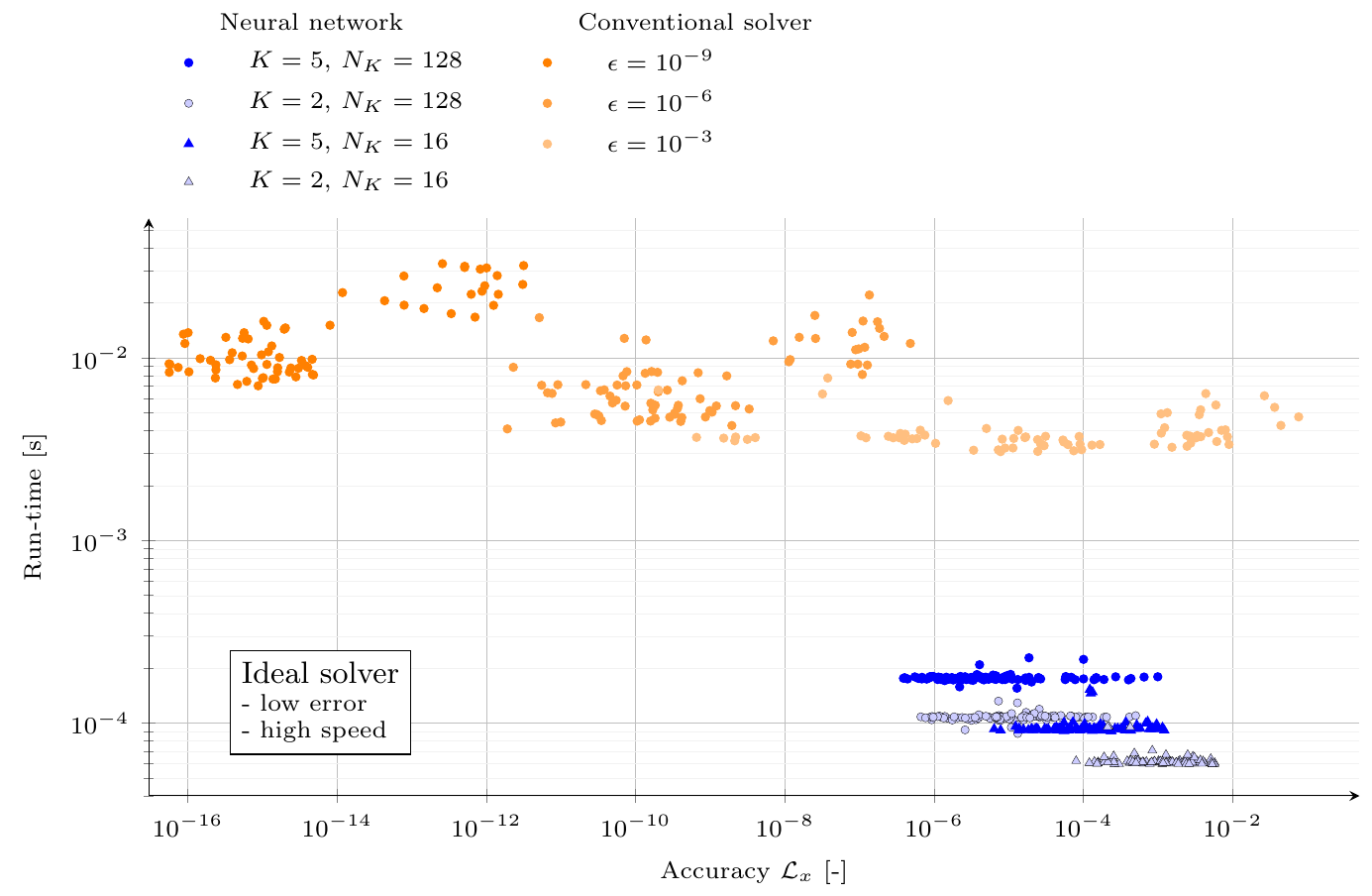}
    \caption{Evaluation of run-time and accuracy for the 11-bus system for varied solver tolerances $\epsilon$ and \gls{NN} sizes ($K$ layers and $N_K$ neurons per layers). Point-wise evaluation for 10 disturbance sizes $P_7$ and prediction time as in \cref{fig:runtime_aspects}.}
    \label{fig:runtime_accuracy}
\end{figure}

\Cref{fig:runtime_accuracy} sets the above results in relation to the achieved accuracy. The points represent different disturbance sizes and prediction times and the accuracy is measured as the associated loss. If a solver yielded points in the lower left corner of the plot, it could be called an ideal solver -- fast and accurate. Conventional solvers can be very accurate when the internal tolerance $\epsilon$ is set low enough, but at the price of being slower to evaluate. Allowing larger tolerances accelerates the solution process slightly at the expense of less accurate solutions. However, this trade-off is limited by the numerical stability of the used scheme; for too high tolerances the results would be considered as non-converged. In case of \glspl{NN}, their superior speed is weighed against less accurate solutions. The accuracy of \glspl{NN} is not only controlled by their size but also, very importantly, by the training process. The achievable accuracy is therefore determined before run-time, in contrast to the tolerance of a conventional solver, which is set at run-time. As a final remark related to \Cref{fig:runtime_accuracy}, we need to highlight that while less adjustable, in contrast to conventional solvers, \glspl{NN} do not face issues of numerical stability as their evaluation is a single and explicit function call.

\begin{figure}[!th]
    \centering
    \includegraphics{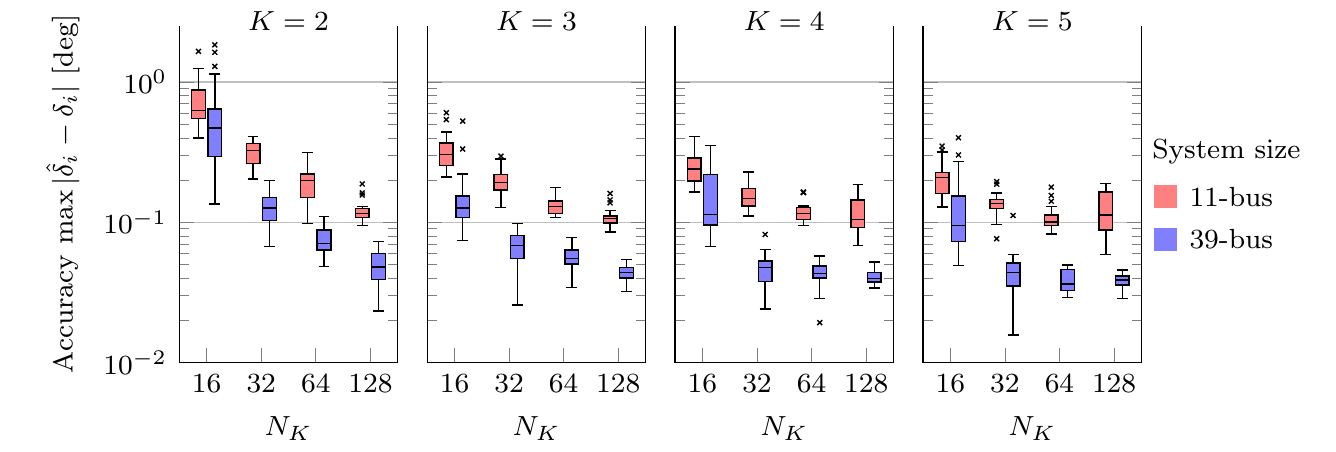}
    \caption{Maximum absolute error of angle $\delta$ on the test dataset for the 11-bus and 39-bus system with varying \gls{NN} sizes, i.e., number of layers $K$ and neurons per layer $N_K$.}
    \label{fig:accuracy_max_errors}
\end{figure}

We lastly want to show how the accuracy, here expressed as the maximum absolute error across all voltage angle states $\max \text{AE}_\delta$ for better intuition, relates to the \gls{NN} size and the power system size. The boxplots in \cref{fig:accuracy_max_errors} represent the evaluation of 20 \glspl{NN} with the same training setup but with different random initialisations of their parameters. We observe that deeper and wider \glspl{NN} usually perform better on this metric. However, the largest \glspl{NN} for the 11-bus system ($N_K=128$ and $K=4$ or $K=5$) show a larger variation than the smaller \gls{NN} which means that the initialisation of the \glspl{NN} affect their performance on the test dataset. This arises in models with a large representational capacity, loosely speaking models with many parameters, hence multiple parameter sets can lead to a low training loss but not all of them generalise well, i.e., have low error on the test dataset. The other, at first sight counter-intuitive, observation is that the 39-bus system performs better on the metric than the smaller 11-bus system. This can be attributed to the complexity of the target function, i.e., of the dynamic responses. The 11-bus system exhibits faster and more intricate dynamics for the presented cases, hence, it is more difficult to approximate their evolution. We could therefore achieve the same level accuracy for the 39-bus system with a smaller \gls{NN} than for the 11-bus system. In terms of run-time, this would mean that the 39-bus system could be faster to evaluate than the 11-bus system. This characteristic of \glspl{NN} effectively overcomes the relationship seen for conventional solvers that larger systems cause longer run-times\footnote{Of course, this relationship breaks when the necessary time step sizes of the conventional solvers differ significantly.} as we have seen in \cref{fig:runtime_aspects}.    

\subsection{\glspl{NN} at training time - a trade-off between accuracy and computational cost}\label{subsec:NN_training}

The benefits of \glspl{NN} compared to conventional solvers at run-time become possible by shifting the computational burden to the \gls{NN} training stage, i.e., the pre-computation of the solution. In this stage, we examine the trade-off between accuracy and the computational cost of the training. This trade-off is influenced by several factors; here, we consider 1) the used training dataset, 2) the type of regularisation, and 3) the optimisation algorithm.
\begin{table}[!th]
    \footnotesize
    \centering
    \renewcommand{\arraystretch}{1.1}
    \caption{Overview of the scenarios with different training datasets.}
    \begin{tabular}{ccccc}
    \toprule
        \thead{Scenario\\\quad} & \thead{Time\\increment $\Delta t$} & \thead{Power disturbance\\increment $\Delta P$} & \thead{Training dataset\\size $|\mathcal{D}_{\text{train}}|$} & \thead{Dataset\\creation cost}\\\midrule
        A & \SI{2}{\second} &\SI{2}{\pu} & 66 & \SI{0.413}{\second}\\
        B & \SI{1}{\second} &\SI{2}{\pu} & 126 & \SI{0.412}{\second}\\
        C & \SI{2}{\second} &\SI{1}{\pu} & 121 & \SI{0.812}{\second}\\
        D & \SI{1}{\second} &\SI{1}{\pu} & 231 & \SI{0.814}{\second}\\
        E & \SI{0.2}{\second} &\SI{0.2}{\pu} & 5151 & \SI{3.880}{\second}\\
        \bottomrule
    \end{tabular}
    \label{tbl:dataset_scenarios}
\end{table}
To investigate the influence of the training dataset and the regularisation, we use the 11-bus system with a \gls{NN} of size $K=5$ and $N_K=32$. We consider five scenarios as shown in \cref{tbl:dataset_scenarios} with different numbers of training data points $|\mathcal{D}_{\text{train}}|$ and the three \say{flavours} of \glspl{NN} which we introduced in \Cref{sec:methodology}: vanilla \gls{NN}, dtNN, \gls{PINN}. The datasets are created by sampling with different increments of time $\Delta t$ and the power disturbance $\Delta P$. As expected, more data points incur a higher dataset creation cost, however, it also depends what \say{kind} of additional data points we generate. When we halve the time increment $\Delta t$, e.g., from scenario A to B or from scenario C to D, the dataset generation cost remains approximately the same. However, this does not hold if we halve the power increment $\Delta P$. When simulating a certain trajectory, it is basically free to evaluate additional points, i.e., reduce $\Delta t$, since interpolation schemes can be used for intermediate points. In contrast, any additional trajectory that needs to be simulated adds to the total cost. Similarly for \say{free}, we can obtain the necessary values for the dtNN regularisation as this only requires the evaluation of the right hand side in \eqref{eq:dynamical_system}. The \gls{PINN} regularisation also incurs only negligible dataset generation cost, as it is a mere sampling of the collocation points $|\mathcal{D}_f|$ without the need for any simulation (here, we use 5151 collocation points that are equally spaced and coincide with the data points from Scenario E). Hence, the additional regularisation comes at no or negligible cost compared to generating more data points unless they lie on trajectories that are evaluated anyways.

\begin{figure}[!ht]
    \centering
    \subfloat[Accuracy on the test dataset \dataset{Test}]{\includegraphics{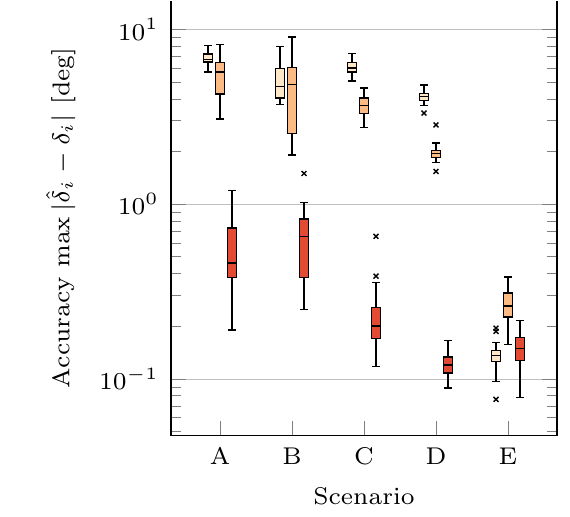}\label{fig:scenario_characteristics_accuracy}}
    \subfloat[Epoch with lowest $\mathcal{L}_x(\dataset{Validation})$]{\includegraphics{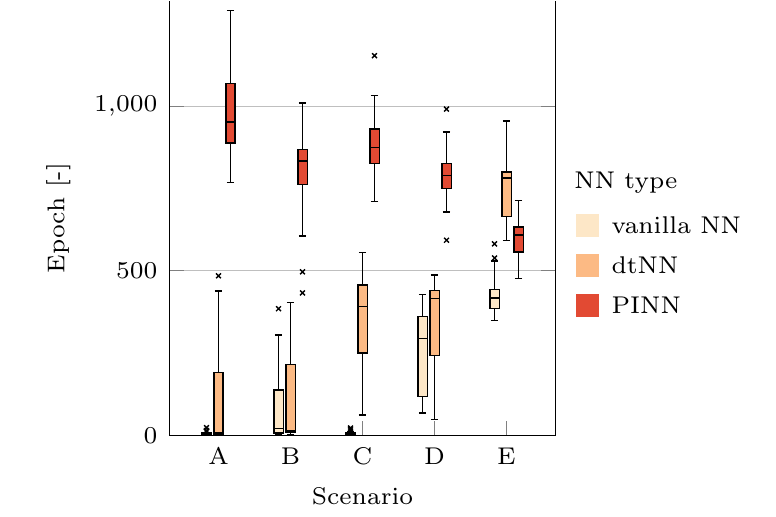}\label{fig:scenario_characteristics_epochs}}
    \caption{Training characteristic for different scenarios (for scenario definition, see \Cref{tbl:dataset_scenarios})}
    \label{fig:scenario_characteristics}
\end{figure}

\begin{figure}[!bh]
    \centering
    \includegraphics{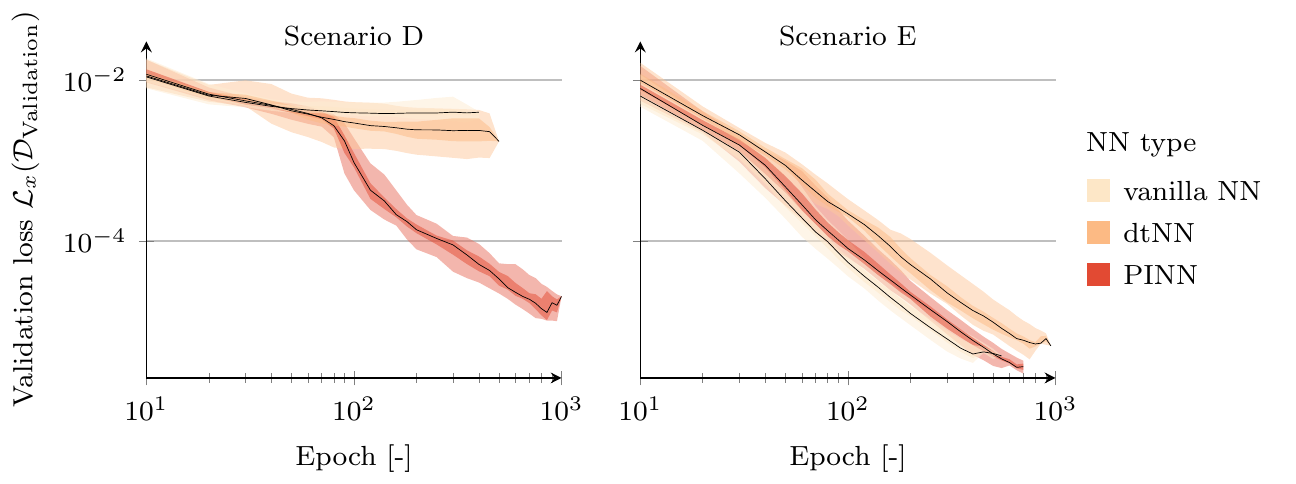}
    \caption{Validation loss as a function of trained epochs. The shadings signify the range from 20 randomly initialised runs.}
    \label{fig:validation_loss_epochs}
\end{figure}

\Cref{fig:scenario_characteristics_accuracy} shows the resulting $\max \text{AE}_\delta$ across 20 training runs with different initialisations of the \gls{NN} parameters. Unsurprisingly, the error metric improves with more data points, i.e., from scenario A to E, and additional regularisation, i.e., from a vanilla \gls{NN} to a dtNN and a \gls{PINN}. In scenario E, which has the largest dataset, all three network types perform on a similar level, whereas \glspl{PINN} otherwise clearly deliver the best performance. Furthermore, the performance becomes more consistent, i.e., less variance, towards scenario E. A very sensitive issue is the point when to stop the training process to prevent over-fitting. In this study, we use the best validation loss as the indicator to determine the \say{best epoch} and \cref{fig:scenario_characteristics_epochs} shows the results. \glspl{PINN} consistently train for more epochs and only for scenario E the three NN types train for approximately the same number of epochs. 
In \cref{fig:validation_loss_epochs} we plot the validation loss over the training epoch. In scenario D, we can clearly see, that while the vanilla \glspl{NN} and the dtNNs do not improve much further after about 100 epochs, \glspl{PINN} still see a significant improvement in terms of accuracy. From around this point onward, the physics-based loss $\mathcal{L}_f$ drives the optimisations, the other training loss terms are already very small. This behaviour partly stems from the fade-in of $\mathcal{L}_f$ but also from the fact that \lossx{} and \lossdt{} are based on much smaller datasets except for scenario E, in which the improvement of the accuracy progresses at similar speeds for all three \gls{NN} types.
\glspl{PINN} offer us therefore the ability to achieve accuracy improvements for more epochs but we can also terminate them early if the achieved accuracy is sufficient to reduce the computational burden. By multiplying the number of epochs with the computational cost per epoch, we can estimate the total computational cost of the training. The vanilla NN and dtNN required about \SI{0.17}{\second} and \SI{0.18}{\second} per epoch for scenarios A-D and \SI{0.40}{\second} and \SI{0.43}{\second} for scenario E while the \gls{PINN} constantly needs \SI{0.66}{\second} per epoch due to the collocation points. These numbers are very implementation and setup dependent, but show the trend that \glspl{PINN} have higher cost per epoch due to the computation of \lossf{} while the dtNN is only slightly more expensive than a vanilla NN. The total up-front cost, comprised of data generation cost and training cost, has then to be evaluated against the desired accuracy to find an efficient setup. This trade-off is again very dependent on the case study. For the 39-bus system the dataset generation cost is 2.5 times more while the cost per epoch only increases by a few percentage points.

\Cref{fig:accuracy_max_errors,fig:scenario_characteristics_accuracy} displayed the maximum absolute errors on the test dataset as the accuracy metric, which is a critical metric of any solution approach. However, the accuracy of a \gls{NN} must also be seen in dependence of the input domain, i.e., the input variables time $t$ and the disturbance size $P_7$.
\begin{figure}[!ht]
    \centering
    \includegraphics{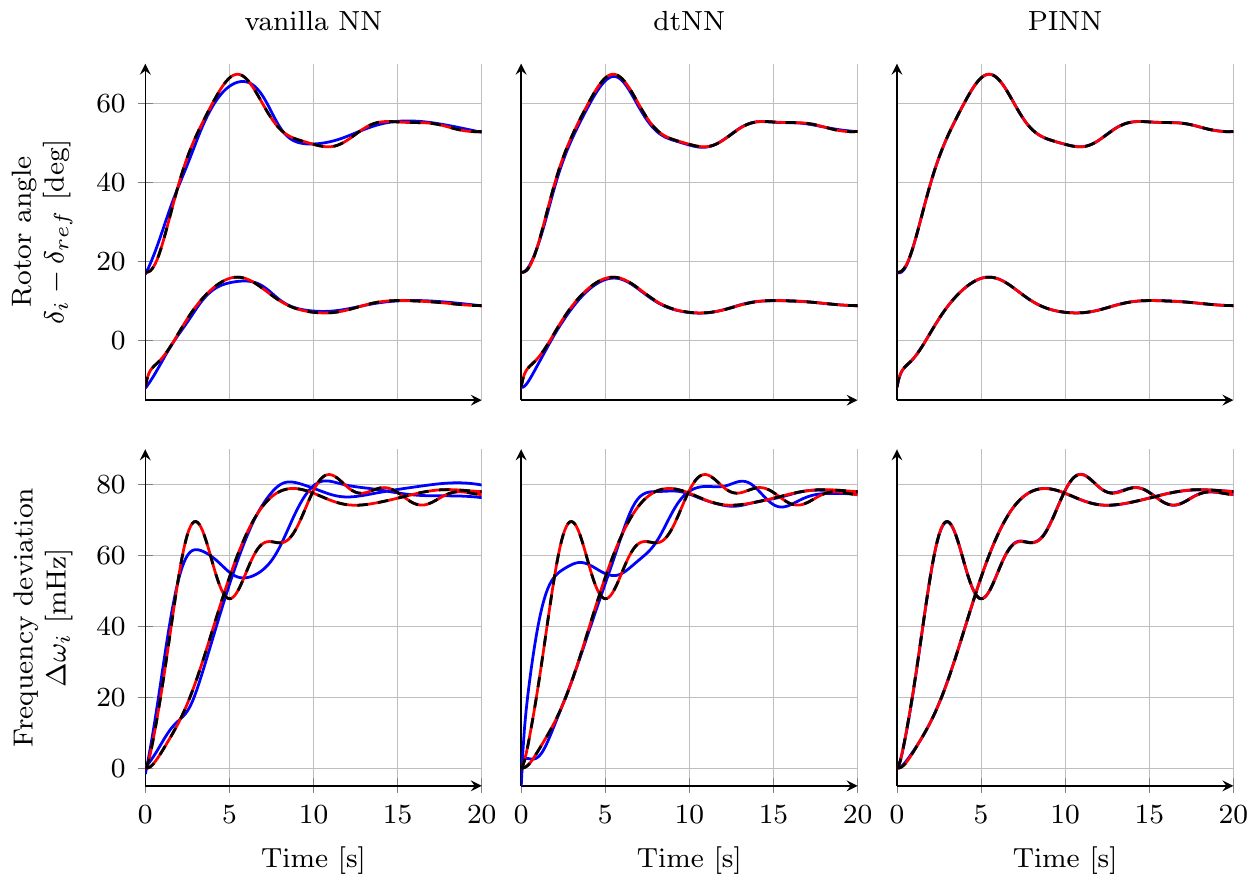}
    \caption{Trajectory of the angle and the frequency deviation at different buses (not the same for $\delta_i$ and $\Delta \omega_i$) based on a training with scenario A (blue) and scenario E (red). Ground truth is shown as the black dashed line and the three columns correspond to the three NN types.}
    \label{fig:trajectories}
\end{figure}
To illustrate this dependence, we show in \cref{fig:trajectories} predictions for the trajectories at two buses resulting from a disturbance of $P_7=\SI{9.1}{\pu}$. The different \gls{NN} types are trained based on scenario A (blue) and E (red) and the black dashed line represents ground truth. In the case of scenario A, NNs and dtNNs still show notable errors, whereas scenario E results in highly accurate predictions. For PINNs, already scenario A yields accurate results, which aligns with the observations in \cref{fig:scenario_characteristics_accuracy}.

\begin{figure}[!th]
    \centering
    \subfloat[Distribution of $\mathcal{L}_x(\dataset{Test})$ as a function of the prediction time $t$]{\includegraphics{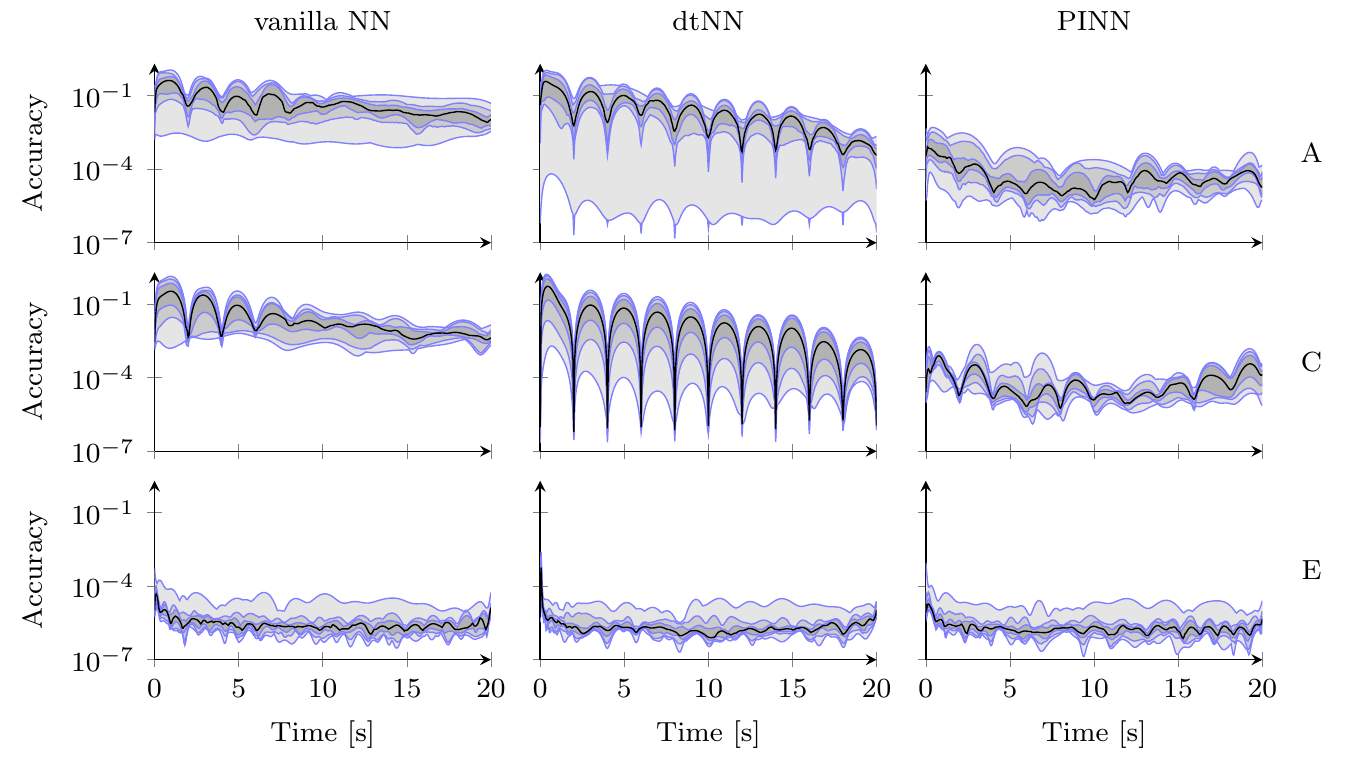} \label{fig:error_characteristics_time}}\\
    \subfloat[Distribution of $\mathcal{L}_x(\dataset{Test})$ as a function of the power disturbance size $P_7$]{\includegraphics{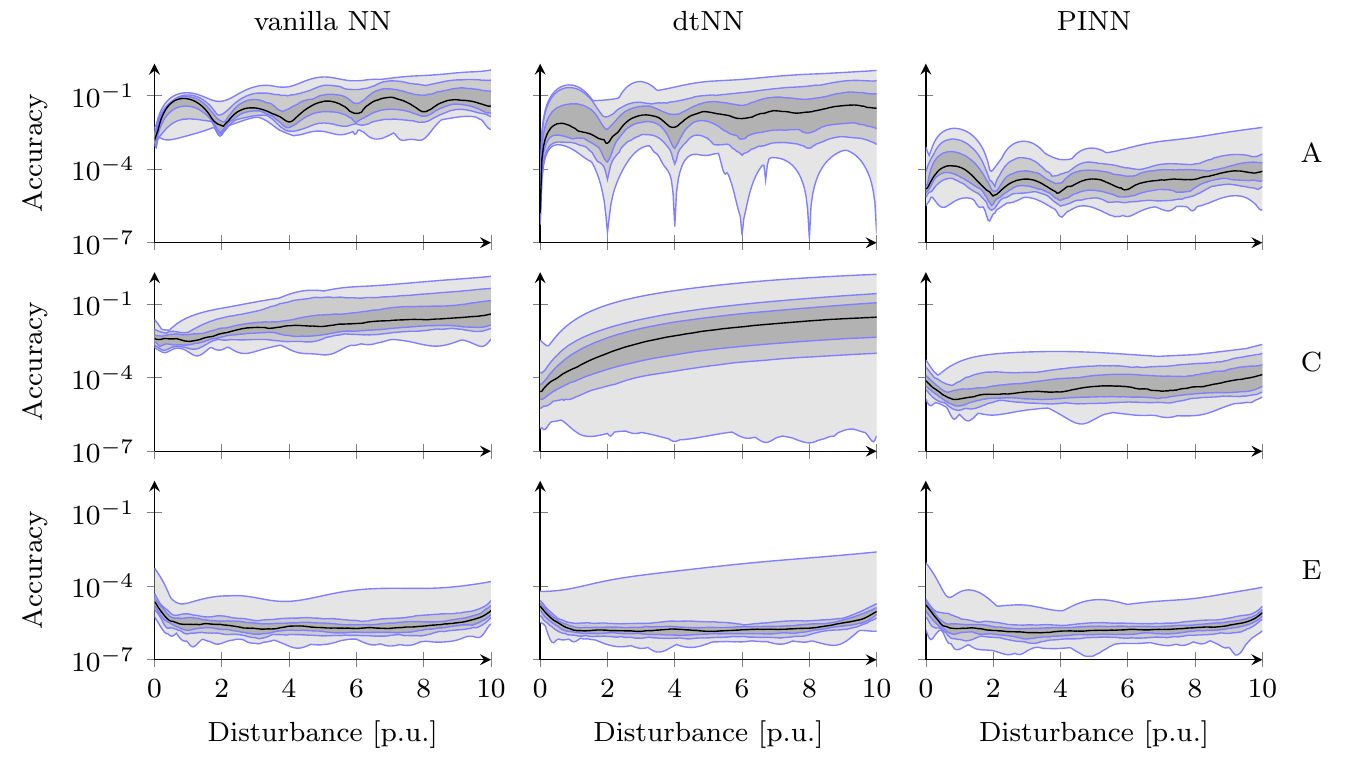}}
    \caption{Distribution of $\mathcal{L}_x(\dataset{Test})$ for different \gls{NN} flavours (vanilla NN, dtNN, PINN) and scenarios (A,C,E). The shaded areas correspond to 100\%, 80\%, 50\% of the errors and the black line represents the median. (for the definition of scenarios A, C, E, see \Cref{tbl:dataset_scenarios})}
    \label{fig:error_characteristics}
\end{figure}

Instead of considering a single disturbance size, we now analyse the prediction performance across the entire test dataset. To this end, we show in \cref{fig:error_characteristics} the resulting distribution of the loss values as a function of the two input variables, i.e., the prediction time $t$ and the disturbance size $\Delta P_i$, for scenarios A, C, and E and the \gls{NN} types. The plots clearly show that for a majority of points in the test dataset, the predictions are much more accurate than the maximum values. This is true in particular around the data points. These are clearly visible in scenario A by the \say{indents}. The panel for the dtNN and scenario C in \cref{fig:error_characteristics_time} shows an extreme case where the prediction at the available data points is very accurate but the interpolation in between produces high errors. In comparison with the vanilla \gls{NN}, the additional regularisation of the dtNN leads to a more unbalanced error distribution. In contrast, the \gls{PINN} shows overall higher levels of accuracy but also more balanced error distributions thanks to the evaluation of the collocation points. We observe two more trends: Smaller prediction times are associated with higher errors due to the faster dynamics; and secondly, larger disturbances tend to show larger errors as they include larger variations of the output variables. These results show the importance of the dataset and the regularisation on the overall characteristics of a \gls{NN}-based predictor, which must be considered for assessing the trade-off between training time and accuracy.

\begin{figure}[!h]
    \centering
    \includegraphics{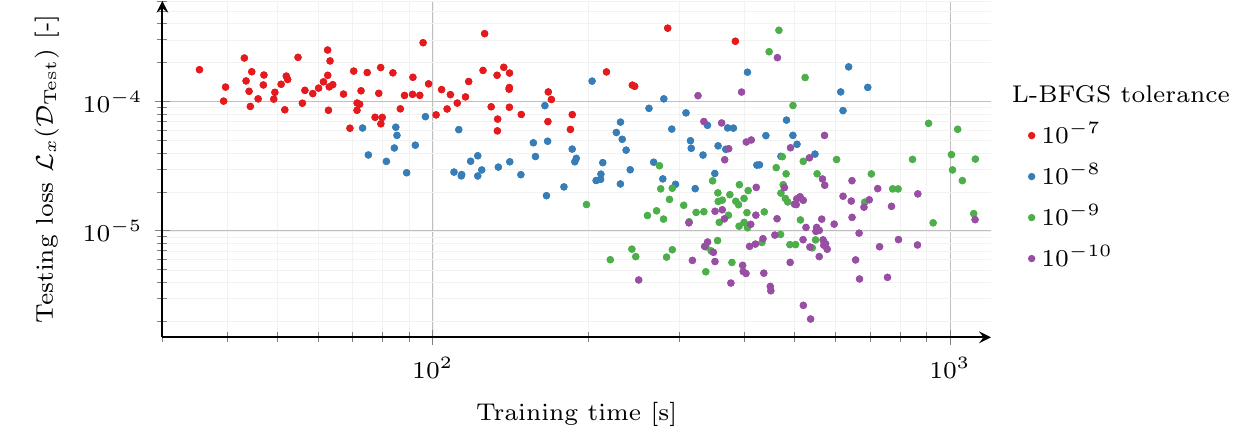}
    \caption{Influence of hyper-parameters of the \gls{LBFGS}-optimser on the trade-off between training time and achieved accuracy. The tolerance level of the optimiser has a large influence as shown by the coloured clusters of points.}
    \label{fig:lbfgs}
\end{figure}

We lastly touch upon the effect of the optimiser on the training process, in this case the \gls{LBFGS} algorithm. The hyper-parameters of the algorithm strongly influence the required training time but also the achieved accuracy which is shown in \cref{fig:lbfgs}. The points represent the outcome from  random hyper-parameter settings and they clearly show a strong relationship between training time and accuracy. Furthermore, the optimiser's internal tolerance setting (coloured) strongly influences this relationship.

\section{Discussion}\label{sec:discussion}
\begin{figure}[!th]
    \centering
    \includegraphics{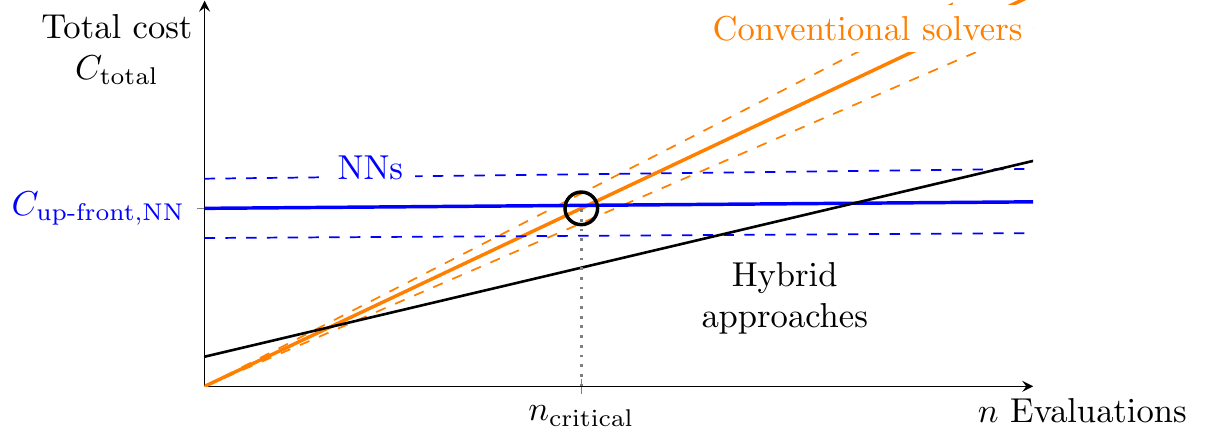}
    \caption{Total cost of different approaches in dependence of the number of evaluations.}
    \label{fig:total_cost}
\end{figure}

The results in the previous Section illustrate how \gls{NN}-based approaches for solving \glspl{DAE} offer a number of advantages at run-time: 10 to 1'000 times faster evaluation speed, no issues of numerical instability, and, in contrast to conventional solvers, their solution time does not increase with a growing power system size. These properties come at the cost of training the \glspl{NN}. To assess an overall benefit in terms of computational time we, therefore, have to consider the total cost as the sum of up-front cost $C_{\text{up-front}}$ for the dataset generation and training and the run-time cost $C_{\text{run-time}}$ per evaluation $n$:
\begin{align}
    C_{\text{total}} = C_{\text{up-front}} + C_{\text{run-time}} \cdot n. \label{eq:total_cost}
\end{align}
\Cref{fig:total_cost} shows a graphical representation of \labelcref{eq:total_cost} for classical solvers and \glspl{NN}. It is clear, that \gls{NN}-based approaches need to pass a critical number of evaluations $n_{\text{critical}}$ to be useful in terms of overall cost, unless other considerations like numerical stability or real-time applicability outweigh the cost consideration. The results in \cref{subsec:NN_run_time,subsec:NN_training} discussed the various \say{settings} that affect run and training time - in \cref{fig:total_cost} they would correspond to the dashed lines. For classical solvers, changing these settings affects the slope, whereas for \glspl{NN} they mostly impact the y-intercept, i.e., $C_{\text{up-front}}$; in either case, as expected, a different \say{setting} will change $n_{\text{critical}}$. Hence, the decision for using \gls{NN}-based methods largely hinges around whether we expect sufficiently many evaluations $n$. Here, it is important to point out that the \gls{NN} will be trained for a specific problem setup and a change in the setup, e.g., another network configuration, requires a new training process. In this aspect of \say{flexibility}, classical solvers have an important advantage over \gls{NN}-based approaches. 

Addressing this lack of flexibility is of paramount importance for adopting \glspl{NN}-based simulation methods and we see three routes forward for this challenge: 1) Reducing the up-front cost $C_{\text{up-front}}$ by tailoring for example the learning algorithms, the used \gls{NN} architectures, and regularisation schemes to the applications; this can largely be seen in the context of actively controlling the trade-off between accuracy and training time. 2) Finding use cases with large $n$, i.e., highly repetitive tasks. 3) Designing hybrid setups -- similar to \gls{SAS}-based methods -- in which repetitive sub-problems are solved by \glspl{NN} and classical solvers handle computations that require a lot of flexibility.

\section{Conclusion}\label{sec:conclusion}
This paper presented a comprehensive analysis of the use of \acrfull{PINN} for power system dynamic simulations. We show that \glspl{PINN} (i) are 10 to 1'000 times faster than conventional solvers, (ii) do not face issues of numerical instability unlike conventional solvers, and, (iii) achieve a decoupling between the power system size and the required solution time. However, \glspl{PINN} are less flexible (i.e. they do not easily handle parameter changes), and require an up-front training cost. Overall, this makes \gls{PINN}-based solutions well-suited for repetitive tasks as well as task where run-time speed is crucial, such as for screening.

Besides the comparison between conventional and \gls{NN}-based methods, this paper conducts a deeper analysis on the parameters that affect the performance of the \gls{NN} solutions. In that respect, we introduce a new \gls{NN} regularisation, called dtNN, as a intermediate step between \glspl{NN} and \glspl{PINN}. We show that \glspl{PINN} achieve overall higher levels of accuracy, and more balanced error distributions thanks to the evaluation of the collocation points.

\section*{Acknowledgement}
The research leading to these results has received funding from the European Research Council under grant agreement no 949899.




 \bibliographystyle{elsarticle-num}
  \bibliography{references_old.bib}





\end{document}